\documentclass[a4paper,12pt]{article}
\usepackage{amsfonts,amsthm,amsmath,amssymb,graphicx,hyperref,youngtab}
\usepackage{bm}
\usepackage{enumerate}
\usepackage{amsmath,amsfonts,dsfont}
\usepackage{amssymb}
\usepackage{tikz}
\usepackage{tikz-cd}
\usetikzlibrary{arrows,decorations.markings}
\usetikzlibrary{calc}
\usepackage[T1]{fontenc}
\usepackage{authblk}
\usepackage[latin9]{inputenc}
\usepackage{color}
\usepackage{latexsym}
\usepackage{amsbsy}
\usepackage{amstext}
\usepackage{graphicx}
\usepackage{amsmath}
\usepackage{amssymb}
\usepackage{tikz}
\usepackage{mathtools}
\usepackage{amsthm}
\usepackage{amstext}
\usepackage{amssymb}
\usepackage{graphicx}
\usepackage{esint}
\usepackage[english]{babel} 

\usepackage{times}
\usepackage{color}
\usepackage{fancyhdr}
\newcommand{\tr}{\operatorname{tr}}

\def\Tr{{\rm Tr\, }}

\newcommand{\cH}{\mathcal H}

\newcommand{\be}{\begin{equation}}
\newcommand{\ee}{\end{equation}}
\newcommand{\bea}{\begin{eqnarray}}
\newcommand{\eea}{\end{eqnarray}}

\DeclareMathOperator{\diag}{diag}
\date{}
\begin{document}
\title{Quiver Asymptotics: $\mathcal{N}=1$ Free Chiral Ring}

\author[1,3 \footnote{E-mail: s.ramgoolam@qmul.ac.uk}]{S. Ramgoolam}
\author[2 \footnote{E-mail: mc.wilson@auckland.ac.nz}]{Mark C. Wilson}
\author[1,3 \footnote{E-mail: a.zahabi@qmul.ac.uk}]{A. Zahabi}
\affil[1]{\footnotesize Centre for Research in String Theory, School of Physics and Astronomy, Queen Mary University of London, UK}
\affil[2]{Department of Computer Science, University of Auckland, Auckland, NZ}
\affil[3]{National Institute for Theoretical Physics,
School of Physics and Mandelstam Institute for Theoretical Physics,
University of the Witwatersrand, Johannesburg, SA}
\maketitle
\abstract

The large $N$ generating functions for the counting of chiral operators in  $\mathcal{N}=1$, four-dimensional quiver gauge theories have previously been  obtained in terms of the weighted adjacency matrix of the quiver diagram. We introduce the methods of multi-variate asymptotic analysis to study this counting in the limit of large charges. 
We describe a Hagedorn phase transition associated with this asymptotics, which refines and 
generalizes known results on  the  2-matrix harmonic oscillator.  Explicit results  are obtained  for two infinite classes of quiver theories, namely the generalized clover quivers and affine $\mathbb{C}^3/\hat{A}_n$ orbifold quivers.

\section{Introduction}
\label{sec:intro}

The AdS/CFT correspondence gives an equivalence between four dimensional gauge theories and ten dimensional string theories \cite{Malda1997}. Generalizations of the correspondence involve four dimensional quiver gauge theories \cite{Do-Mo} and a six-dimensional non-compact Calabi-Yau  (CY) space in the transverse directions.  The dictionary between $\mathcal{N}=1$ four-dimensional   quiver gauge theories and the Calabi-Yau geometry has been developed, in the case of toric CY \cite{Ful},  using brane tilings  \cite{Hanany:2005ve,Franco:2005rj,Ken}. 

Chiral gauge invariant operators, which are annihilated  by supercharges of one chirality, play a central role in identifying the CY  space for a given quiver gauge theory. These operators form a chiral ring
and their expectation values are independent of the positions of insertion  of the operators \cite{CDSW}. 

The combinatorics of the chiral ring at non-zero super-potential  has been studied using generating functions and  Hilbert series 
in the plethystic program \cite{Fe-Ha-He,Be-Fe-Ha-He,Ha-Za}. 
The asymptotics of the counting formulae have also been studied \cite{Do, Be-Fe-Ha-He, Fe-Ha-He,Lucietti:2008cv}. These studies have  primarily used one-variable methods appropriate for special cases of the asymptotics, although a few results in the multi-variable case are also available \cite{Do,Lucietti:2008cv}.

General results on the counting of chiral ring operators in the free limit of zero superpotential 
were obtained  in \cite{Pa-Ra,Ma-Ra}. The generating function for chiral operators in this free limit is  an infinite product of inverse determinants involving the weighted adjacency matrix of the quiver diagram. 
This weighted adjacency matrix is a function of multiple variables associated with the edges of the quiver, and corresponding charges. The asymptotic behaviour of the free quiver counting in the limit of large charges naturally requires multi-variate complex analysis.  
In this paper, we will study the asymptotics of  these generating functions using the methods of multi-variate asymptotics recently developed by Pemantle and Wilson \cite{Pe-Wi, Pe-Wi-book}. These methods allow systematic algorithmic derivation of asymptotics associated to a rational generating function in several variables, and apply directly to the generating function of chiral operators.

We find that the asymptotic counting of the chiral operators for any free quiver gauge theory is given by a compact general  formula~\eqref{asymp form}. 
To make this general formula more explicit  for different quivers requires, at present, symbolic computer tools. For two infinite classes of examples, we analytically  derive the explicit asymptotic results. 
These two classes are  the generalized clover quivers and the affine $\mathbb{C}^3/\hat{A}_n$ orbifold quivers. These  results are the asymptotic formulas \eqref{free asympt} and \eqref{free asympt 3}.

The rest of the paper is organized as follows. In Section~\ref{sec:quiver}, we define  a  thermodynamics of 
the chiral operators in a free quiver gauge theory, based on the counting problem of 
the chiral operators. We discuss associated phase transitions and a generalized Hagedorn hyper-surface related to the asymptotic analysis.  In Section~\ref{sec:asymp}, we present an adapted version of multivariate asymptotic techniques for our quiver gauge theory problem. In Section~\ref{sec:classes}, we apply the developed method to two (infinite) classes of examples and obtain explicit results for the  asymptotics. Section~\ref{sec:future} discusses some possible directions for future studies.

\section{Thermodynamics of Chiral Ring in Free CY$3$ Quiver Theories}
\label{sec:quiver}

In this section, we will consider generating  functions for the counting of chiral ring operators in free quiver gauge theories as generalizations of thermal partition functions in AdS/CFT. We will  define a generalized Hagedorn hyper-surface. We will observe that it  controls the asymptotics of these generating functions, a point which will be developed in more detail in subsequent sections. 

In the context of the Anti-de-Sitter/Conformal Field Theory (AdS/CFT)  correspondence,  type IIB string theory on $AdS_5 \times S^5$ is dual  to $\mathcal{N}=4$ Super Yang-Mills (SYM) theory on four dimensional Minkowski space $\mathbb{R}^{1,3}$.  Conformal field theories have a symmetry of scaling of  the space-time coordinates. Their quantum states are characterized by a scaling dimension, which is the eigenvalue of 
a scaling operator. In the AdS/CFT correspondence, the scaling  operator of the CFT corresponds to
the Hamiltonian for global time translations in AdS \cite{Wit1}. The eigenvalues of this Hamiltonian 
are the energies of quantum states obtained from the quantum theory of gravity in Anti-de-Sitter space.

For free CFTs in four space-time dimensions, the scaling dimension for any  scalar field is 1. 
For a composite field (also called composite operator), which is a monomial function of  the elementary scalar fields, the scaling dimension is the number of constituent scalar fields. 

The thermal partition function of the AdS theory is a function of $\beta $, the inverse temperature, given by 
 \bea 
 F ( \beta ) = \Tr e^{ - \beta H }. 
 \eea
For a system with a discrete spectrum of energies, as in the  case at hand, this is a sum over energy eigenvalues 
\bea 
F(\beta)=\sum_E ~ a (E) ~  e^{-\beta E},
\eea 
where $ a ( E ) $ is the number of states of energy $E$. 
For a free CFT, the partition function becomes 
\bea 
\label{single part func}
F ( x ) = \sum_r ~  a_r ~  x^r,
\eea
where $ x = e^{ - \beta } $,  $a_r$ is the number of composite fields with $r$ constituent 
elementary scalars and $r$ is being summed over the natural numbers. In a case 
where we have multiple types of scalar fields, as in quiver gauge theories, the above 
partition function can be generalized to a multi-variable function  
\bea 
\label{multi part func}
F ( x_1 , x_2 , \cdots ) = \sum_{ r_1 , r_2 , \cdots }  ~ a_{ r_1 , r_2 , \cdots } ~  x_1^{ r_1} x_2^{ r_2} \cdots ,
\eea
where $x_i= e^{-\beta_i}$ and $ a_{ r_1 , r_2 , \cdots } $ is the multiplicity of composite operators with specified numbers 
$r_1 , r_2 , \cdots $ of the different types of scalar fields. The multivariate generating functions (\ref{multi part func}) are refined versions of the partition function (\ref{single part func}). 
We will refer to $r_i$ as charges and the  variables  $x_i$  as fugacity factors for each field. It is convenient to introduce the following vector notation $\bm{\beta}=(\beta_1, \beta_2, \cdots)$, $\textbf{r}=(r_1, r_2, \cdots)$, ($\textbf{r}=|\textbf{r}|\hat{\textbf{r}}$, $|\textbf{r}|=\sqrt{r_1^2+r_2^2+ \cdots}$) and  define 
\bea 
\mathbf{x}^{\mathbf{r}}  :=  \prod_{j=1}^{d}x_j^{r_j}.
\eea 
We can then write Eq. (\ref{multi part func}) as
\begin{equation}
\label{multi part func 2}
F(\textbf{x}) =\sum_{\textbf{r}}a_{\textbf{r}} ~ \textbf{x}^{\textbf{r}}.
\end{equation}

The 1-variable partition function (\ref{single part func}),  for systems which have 
an exponential growth of  the number of states  $a_r \sim e^{\alpha r}$ in the large $r$ limit, 
has regions of convergence and divergence meeting at a critical $ \beta = \alpha$. 
The partition function converges for $ \beta   > \alpha $ and diverges for $ \beta < \alpha$. 
This type of behaviour  occurs in string theory, where it
 is associated with the Hagedorn phase transition \cite{Gr-Wi-Sc}. 
 
Similarly, in the multi-variable case, assuming that the function $ F ( x_1 , x_2 , \cdots  )  $  has an exponential growth of the multiplicity factor 
$ a_{ r_1 , r_2 , \cdots } $ for large $r_i$, then there is a hyper-surface separating 
convergence and divergence regions for the multi-variable partition function. This hyper-surface 
is given by the equation 
\bea 
\label{Hag type}
\frac{1}{|\textbf{r}|}\log a_{\textbf{r}} = \bm{\beta}\cdot \hat{\textbf{r}}\quad .
\eea 
In the quiver examples we will be studying in this paper, there is indeed this type of 
exponential behaviour and  a corresponding hyper-surface. This may be viewed as 
 a generalized Hagedorn hyper-surface.

This Hagedorn hyper-surface was studied in the case of a 2-matrix model in \cite{Aha}. The 2-matrix model problem is associated with a quiver consisting of a single node and 2-directed edges. We will revisit 
this model and consider the generalised  Hagedorn hyper-surface for  more general partition functions associated with quivers \cite{Pa-Ra}.

A \emph{quiver diagram} is a directed graph $G=(V,E)$ with a set $V$ of \textit{nodes} and a set $E$ of directed \emph{edges}; self-loops at a vertex are explicitly allowed. A quiver gauge theory has a  product 
gauge group of the form  $\prod_a U(N_a)$, where each $U(N_a)$ is associated with a node, and matter fields in the bi-fundamental representation of the gauge group are associated with the edges. The interactions between the matter fields are described by a \textit{superpotential $W$} which is a gauge-invariant polynomial in the matter fields. In our study, we focus on the \textit{zero superpotential} case $W=0$.

The observables of  quiver gauge theories are \textit{gauge invariant operators} and their correlation functions. An interesting class of observables is formed by  chiral operators, which form a ring, called the chiral ring. For the definition and properties of the chiral ring, see for example \cite{CDSW,Ha-Za}. 
The space of chiral operators in the zero-superpotential limit is typically much larger than that  at non-zero superpotential.

The generating function for the chiral operators in the large $N$ limit in an arbitrary free (with zero superpotential $W=0$) quiver gauge theory was derived in  \cite{McG,Pa-Ra,Ma-Ra} and is given by 
\begin{equation}
\label{free arbit quiv}
F(\mathbf{x}) =\sum_{\mathbf{r}}a_{\mathbf{r}}\mathbf{x}^{\mathbf{r}}= \prod_{i=1}^\infty \det(\mathbb{I} - X(\mathbf{x}^i))^{-1}, 
\end{equation}
where we have introduced  fugacity factors 
$\mathbf{x}=(x_1, x_2, ..., x_d)$ and  charges $\mathbf{r}=(r_1, ..., r_d)$ associated with  quiver edges. 
The  weighted adjacency matrix of the quiver diagram  has been denoted by $X(\mathbf{x}) = 
X ( x_1 , x_2 , \cdots , x_d )$ and $ X(\mathbf{x}^i ) = X ( x_1^i , x_2^i  , \cdots , x_d^i )$. 
 The multivariate generating function \eqref{free arbit quiv} is an example of Eq.~\eqref{multi part func 2}, i.e. a refined version of the partition function \eqref{single part func}. The \emph{ degeneracy } $a_{\mathbf{r}}$ is the number of chiral operators with \emph{charge vector} $\mathbf{r}$ in the chiral ring of the free quiver gauge theory. In this paper, we will be interested in 
the asymptotic behaviour of $ a_{ \mathbf{r} } $ for large $( r_1 , r_2 , \cdots , r_d ) $. 

In order to study this asymptotics, with the methods of  \cite{Pe}, \cite{Pe-Wi} and \cite{Pe-Wi-book}, it will be useful to write $H$ as a ratio 
 $$F(\mathbf{x})=\prod_{i=1}^\infty \det(\mathbb{I} - X(\mathbf{x}^i))^{-1}=\frac{G(\mathbf{x})}{H(\mathbf{x})},$$ 
where 
\bea
\label{H}
H(\mathbf{x})= \det(\mathbb{I} - X(\mathbf{x})),\quad  G(\mathbf{x})=\prod_{i=2}^\infty \det(\mathbb{I} - X(\mathbf{x}^i))^{-1}.
\eea
In fact, we observe that, with the parameterization $\mathbf{x}= \exp{(-\bm{\beta})}$, $F(\mathbf{x})$ is convergent in the domain $H(\mathbf{x})\geq0$ for small enough (and positive) $x_i>0$ for all $i$ . The boundary between the convergence and divergence domains is the phase transition hyper-surface, characterized by $H(\mathbf{x})=0$. We will see some examples of this in Section~\ref{sec:classes}.  The  asymptotic regime of the generating function is obtained by approaching the $H=0$ hyper-surface inside the domain of convergence. Thus this hyper-surface  controls the phase structure of the theory and also  determines the leading asymptotic behavior of $a_{\mathbf{r}}$. 
%Dominance of $H=0$ term in the asymptotic behavior of the degeneracy $a_{\mathbf{r}}$ is based on partial analytic and numerical computations, see Appendix \ref{App B}.
The asymptotic analysis will be developed in Sections ~\ref{sec:asymp} and \ref{sec:classes}.

\subsubsection*{Entropy}
The logarithm of the degeneracy $ a_{\mathbf{r}  } $ is the {\it thermodynamic entropy } 
$S_{\mathbf{r}}\coloneqq \log a_{\mathbf{r}}$. For convenience, sometimes we consider the leading term of the entropy, which we denote by $S_{\mathbf{r}}^*$.

Following general result \eqref{Hag type}, in the chiral ring of the free quiver gauge theories, the Hagedorn-type transition can be seen as a result of the competition between the leading term of the entropy $S^*_{\mathbf{r}}$ obtained from the logarithm of the multiplicity and the temperature term $-\bm{\beta}\cdot \mathbf{r}$ in the generating function \eqref{free arbit quiv}. 
The  generalized Hagedorn hyper-surface is given by 
\bea
\label{Legendre}
S^*_{\mathbf{r}}-\bm{\beta}\cdot \mathbf{r}=0.
\eea
Using Eq.~\eqref{Legendre}, the critical couplings can be simply obtained as $\beta_i=\partial _i S^*_{\mathbf{r}}$. We will see explicit equations for this hyper-surface 
 in some classes of examples in Section~\ref{sec:classes}.
 
\section{Method of Asymptotic Analysis}
\label{sec:asymp}

In this part we adopt a novel technique of asymptotic analysis of the multivariate generating functions, and apply it to the counting problem for the corresponding quiver gauge theories. First, we review some known material from the asymptotic analysis of multivariate generating functions and then in the second part, we present an ongoing development on the evaluation of some Hessian determinant specified at some critical points. In the third part, the phase structure of the quiver theories is explained and the relations between the entropy and critical couplings are discussed.

\subsection*{Multivariate Asymptotic Counting}

In this section, we answer the question of asymptotic counting for the multivariate generating functions that appear in the chiral ring of the quiver gauge theories.
We will not review the details of the proofs from multivariate asymptotic analysis in this article and only present the main result in the following. For a comprehensive presentation of such analysis see \cite{Pe}, \cite{Pe-Wi} and \cite{Pe-Wi-book}.

 %\subsubsection*{$\bullet$ Multivariate asymptotics}
We now briefly summarize the general results for asymptotics of multivariate generating functions obtained by Pemantle and Wilson \cite{Pe-Wi}, applied to our situation of interest. In the present paper, we encounter generating functions that only require the \emph{smooth point} analysis of \cite{Pe-Wi}. We  present an adapted and extended version of these results, in four steps suitable for quiver gauge theories. In the next section we apply these results to some infinite classes of examples of quivers. 

The basic steps of the analysis of \cite{Pe-Wi} are as follows.
\begin{enumerate}[(i)]
\item We consider a generating function in $d$ variables $\mathbf{x}=(x_1, ..., x_d)$, 
\bea 
F(\mathbf{x})=\frac{G(\mathbf{x})}{H(\mathbf{x})}.
\eea
where $G$ and $H$ are holomorphic in some neighbourhood of the origin, $H(\mathbf{0}) \neq 0$, and all 
coefficients are nonnegative.
\item For a given $\mathbf{r}$ we find the \emph{contributing points} to the asymptotics in direction parallel to $\mathbf{r}$. To do this we first find the \emph{critical points} of $H$. A smooth critical point $\mathbf{x}^*$ is a solution of the following set of equations
\bea 
\label{Critical Eqs}
H(\mathbf{x}^*)=0, \quad  r_d x^*_j \partial_j H(\mathbf{x}^*)=r_j x^*_d\partial_d H(\mathbf{x}^*)\qquad  \text{for}\ (1\leq j\leq d-1).
\eea
From general theory \cite{Pe-Wi-book} there is a  solution $\mathbf{x}^*$ to these equations that has only positive coordinates and which is a contributing point. Generically, this is the unique solution to the critical point equations. Furthermore all other contributing points, if they exist, lie on the same torus. We require $F = G/H$ to be meromorphic in a neighbourhood of the \emph{closed polydisk} containing $\mathbf{x^*}$, defined by the condition that $|x_i| \leq x_i^*$ for all $i$. The point is usually \emph{strictly minimal} --- no other point in the polydisk is a pole of $F$.

\item Near each contributing point $\mathbf{x^*}$ we can solve the equation $H(\mathbf{x}^*)=0$, without loss of generality $x_d=g(\mathbf{x}')$ with $\mathbf{x}'=(x_1, ..., x_{d-1})$. However, the asymptotic is independent of which coordinate we solve for. Then we define the function $\phi=\log g(\mathbf{x}')$ locally parametrizing the hyper-surface $\{H=0\}$ in logarithmic coordinates. The Hessian $\mathcal{H}$ of $\phi$ is an essential part of the asymptotic formula, and we can compute it in terms of the original data as follows. 
We construct the Hessian matrix $\mathcal{H}\coloneqq \Big(\mathcal{H}_{ij}\Big)_{i,j=1}^{d-1}$ with elements $\mathcal{H}_{ij}=\frac{\partial^2{\phi}}{\partial x_i\partial x_j}$. The diagonal and off-diagonal matrix elements can be written  explicitly in terms of $g$:
\bea 
\label{elements of Hessian}
\mathcal{H}_{ii}&=&-x_i\frac{\partial_{i}g(\mathbf{x}')+x_i\partial^2_{i}g(\mathbf{x}')}{g(\mathbf{x}')}+ x_i^2 \Big(\frac{\partial_{i}g(\mathbf{x}')}{g(\mathbf{x}')}\Big)^2 \ \quad\text{diagonal},\nonumber\\
\mathcal{H}_{ij}&=&-x_i x_j \Big(\frac{\partial_{i} \partial_{j}g(\mathbf{x}')}{g(\mathbf{x}')}- \frac{\partial_{j}g(\mathbf{x}')\partial_{i}g(\mathbf{x}')}{g(\mathbf{x}')^2}\Big)\ \quad \text{off-diagonal}.
\eea
Thus we can write the Hessian matrix as
\bea
\label{Hessian 1}
\mathcal{H}=\left(\frac{x_i x_j}{g}\left(\frac{\partial_{i}g\partial_{j}g}{g}-\partial_{i}\partial_{j}g\right)- \frac{x_i\partial_{i}g}{g}\delta_{ij}\right)_{i,j=1}^{d-1}.
\eea
The above holds for all values of the variables. We are only interested in the evaluation of the Hessian determinant at each contributing point, and this allows further simplification which we now carry out. The critical equation $H=0$ implies
\bea
 \quad dH= \sum_{i=1}^{d-1} \partial_i H\,  dx_i+ \partial_d H\, dx_d=\sum_{i=1}^{d-1} \left[\partial_i H + \partial_d H\, \frac{\partial x_d}{\partial x_i}\right] dx_i=0,
\eea
where we omitted the star for the critical points for simplicity.
It implies the following relation  for any $j=1, ..., d-1$: 
\bea
\frac{\partial_j H}{\partial_d H}= - \partial_j g.
\eea
Putting this relation together with \eqref{Critical Eqs}, 
we obtain
\bea
\label{critical ratio}
\frac{x_j \partial_j g}{x_d}= -\frac{r_j}{r_d}.
\eea

Applying identity \eqref{critical ratio} to the Hessian matrix \eqref{Hessian 1} we obtain the elements of the Hessian matrix evaluated at critical points,
\bea
\label{Hessian 2}
\mathcal{H}_{ij}^*=\frac{r_i r_j}{r_d^2}-\frac{x_i^* x_j^*\partial_{i}\partial_{j}g}{g}+ \frac{r_i}{r_d}\delta_{ij}.
\eea
The computation of $\frac{x_i^* x_j^*\partial_{i}\partial_{j}g}{g}$ depends on the relative positions of the loops $i$, $j$ and $d$ in the quiver diagram.
\item  The final step is to derive the asymptotic formula for $a_\mathbf{r}$ from the critical points and Hessian determinant. The Cauchy integral formula yields
\bea
a_{\mathbf{r}}=\frac{1}{(2\pi \mathrm{i})^d}\int_T \mathbf{x}^{-\mathbf{r}} F(\mathbf{x}) \frac{d \mathbf{x}}{\mathbf{x}},
\eea
where the torus $T$ is a product of small circles around the origin in each coordinate and $\frac{d \mathbf{x}}{\mathbf{x}}= (x_1 \cdots x_d)^{-1} dx_1 \wedge \cdots \wedge dx_d$.
In the asymptotic regime $|\mathbf{r}|\rightarrow \infty$, the following \emph{smooth point asymptotic formula} is obtained by Pemantle and Wilson, see Theorem (1.3) in \cite{Pe-Wi}. We write simply $\mathbf{x}$ for $\mathbf{x}^*(\mathbf{r})$. 

The smooth point formula states that if $G(\mathbf{x})\ne 0$ then
\bea
\label{asymp form}
a_{\mathbf{r}}\sim (2\pi)^{-(d-1)/2}(\det\mathcal{H(\mathbf{x})})^{-1/2}\frac{G(\mathbf{x})}{-x_d \partial H/\partial x_d(\mathbf{x})}r_d^{-(d-1)/2}\mathbf{x}^{-\mathbf{r}},
\eea
where $\mathbf{x}^{-\mathbf{r}}\coloneqq\prod_{j=1}^{d}x_j^{-r_j}$. The expansion is uniform in the direction $\hat{\mathbf{r}}\coloneqq\mathbf{r}/|\mathbf{r}|$ provided this direction is bounded away from the coordinate axes.
\end{enumerate}

We want to apply the above procedure in our case of interest. For any connected quiver with generating function \eqref{free arbit quiv}, we have functions $H(\mathbf{x})$ and $G(\mathbf{x})$ as in Eq.~\eqref{H}.
For $\mathbf{x}$ sufficiently close to the origin, $G$ and $H$ are holomorphic and $H$ does not vanish at the origin.

Our next observation is the rediscovery of a folklore result in graph theory \cite{Cve}, that $H(\mathbf{x})$ in Eq.~\eqref{H} can be expanded graphically in terms of the loops in the quiver,
\bea
\label{loop exp H}
H= \sum_{k=0}^d\sum_{l_1\sqcup l_2\sqcup ... \sqcup l_k} (-1)^k l_1 l_2 \cdots l_k,
\eea
where $l_1$, $l_2$, ..., $l_k$ are the loops which meet each node of the quiver diagram only once, $d$ is the total number of the loops of the quiver diagram, and loops in the second sum are disjoint. Notice that each loop variable in the above expansion \eqref{loop exp H} is the product of $x_i$ edge variables in the determinant formula \eqref{H}, around each loop of the quiver diagram.

To summarize, given a quiver diagram, one can easily find the function $H$ and solve the critical equations to obtain the critical points. Then, by computing the Hessian determinant evaluated at critical point and inserting these results into Eq.~\eqref{asymp form}, one can obtain the asymptotic for any quiver diagram. However, owing to the multidimensional nature of the problem, some of the computations, such as solving the critical equations and computing the Hessian determinant, require symbolic mathematical software. On the other hand, as we will present in this paper, alternatively, one can try to find an explicit analytic form of the asymptotic formula for some infinite classes of quivers, with the hope of finding a general analytic result for larger classes of quivers.  

\section{Some Infinite Classes of Quivers}
\label{sec:classes}

Having introduced and discussed a general procedure of the asymptotic methods for quiver diagrams, in this section, we implement these methods in two infinite classes of examples and obtain explicit analytic results for the entropy and phase structure of these quiver gauge theories. 
\subsection{Generalized Clover Quivers}
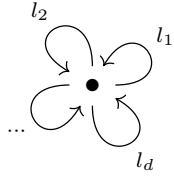
\begin{figure}
\centering
\begin{tikzcd}
\bullet \arrow[out=0,in=60,loop,swap,"l_1"]
  \arrow[out=90,in=150,loop,swap,"l_2"]
  \arrow[out=180,in=240,loop,swap,"\cdots"]
  \arrow[out=270,in=330,loop,swap,"l_d"]
\end{tikzcd}
\caption{Generalized Clover Quiver}
\label{clover}
\end{figure}

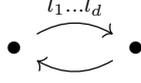
\begin{figure}
\centering
\begin{tikzcd}
\bullet \arrow[r,bend left,"l_{1}...l_{d}"] & \bullet \arrow[l,bend left] 
\end{tikzcd}
\caption{$d$-Kronecker Quiver}
\label{kronecker}
\end{figure}

As the first class of examples, we consider a generating function of the following form:
\bea
\label{free}
F(\mathbf{x}) = \frac{G(\mathbf{x})}{H(\mathbf{x})}=\sum_{\mathbf{r}} a_{\mathbf{r}} \mathbf{x}^\mathbf{r}= \prod_{i=1}^\infty \left(1  -\sum_{j=1}^{d}x_j^i\right)^{-1},
\eea
with $H(\mathbf{x})=1-\sum_{j=1}^{d} x_j$ and $G(\mathbf{x})=\prod_{k\geq 2} H_k(\mathbf{x})=\prod_{i=2}^\infty (1  -\sum_{j=1}^{d}x_j^i)^{-1}$. This is the \textit{Generalized Clover Quiver} class, see Fig.~\ref{clover}. It is interesting to notice that the $d$-Kronecker quivers,  consisting of $d$ loops, shown in Fig.~\ref{kronecker}, have also the same generating function.

It is easy to observe that for direction $(r_1, \dots, r_d)$ the unique critical point is $\mathbf{x}^*=(\frac{r_1}{R}, \frac{r_2}{R}, ..., \frac{r_d}{R})$, with $R=\sum_{i=1}^{d}r_i$.  This point is strictly minimal. To see this, we first show that there are no more zeros of $H_1$ in the closed polydisk defined by $\mathbf{x^*}$. This is clear because for every such point $\mathbf{x}$  we must have $\sum_{i=1}^d x_i = 1$ while $|x_i|\leq x_i^*$, and this can only happen when 
all $x_i = x_i^*$. We then handle possible poles arising from the other factors $H_k$ for $k\geq 2$.
To do this, note that $0 < x_i^* < 1$ for all $i$. Thus if $|x_i|\leq x_i^*$ for all $i$ and $k\geq 2$ we have

\begin{equation*}
\left| 1 - \sum_{i=1}^d x_i^k \right|  \geq 1 - \sum_{i=1}^d |x_i|^k \geq 1 - \sum_{i=1}^d |x_i| 
 \geq  1 - \sum_{i=1}^d x_i^* = 0.
\end{equation*}
Note that the second inequality is strict unless all $x_i = 0$, in which case the third inequality is definitely strict. Thus  we can choose a polydisk with radii $f_i$ slightly more than $x_i^*$, such that for all $k\geq 2$,  $\sum_i f_i^k < 1$ and so each factor in the product defining $G$ is analytic. Thus $F = G/H$ has the desired form as a quotient of analytic functions in an appropriate polydisk, and $\mathbf{x^*}$ is a strictly minimal critical point for the given direction.

To compute the Hessian, we observe that $g=x_d= 1- \sum_{i=1}^{d-1} x_i$ is linear and so the second partial derivatives are zero. Thus from \eqref{Hessian 2} we have 
$$
\mathcal{H}_{ij} = \frac{r_ir_j}{r_d^2} + \frac{r_i}{r_d}\delta_{ij}.
$$
By changing to the variables $r_i/r_d$ the determinant of this matrix is easily computed (see Appendix \ref{App A}) to be 

\bea
\label{Clover det}
\det\mathcal{H}=r_d^{-(1+d)}R \prod_{i=1}^{d}r_i.
\eea

By implementing the above explicit formula for the determinant of the Hessian matrix, in Eq.~\ref{asymp form}, it is straightforward to obtain the asymptotic result. By using Eq.~\eqref{asymp form}, for the asymptotics of $a_{\mathbf{r}}$ of the generating function \eqref{free}, in the "central region",  as $R \to \infty$ and $r_i/r_j$ (for each $i,j=1\ \text{to}\ d$) bounded away from zero, we obtain the following explicit asymptotic formula,
\bea
\label{free asympt}
a_{\mathbf{r}}\sim\frac{G(\mathbf{x}^*)} {(2\pi)^{(d-1)/2}}R^{R+\frac{1}{2}}\prod_{i=1}^{d}r_i^{-r_i-\frac{1}{2}}.
\eea 

The entropy of the generalized clover quiver is obtained in the following Shannon form:
\bea
\label{entropy clover}
S_{\mathbf{r}}= \log G(\mathbf{x}^*)+ \left(R+\frac{1}{2}\right)\log R-\sum_i \left(r_i+\frac{1}{2}\right) \log r_i.
\eea

\subsubsection*{$\bullet$ Univariate Case }
In this part we characterize a special type of quiver whose asymptotics cannot be studied with the methods above. This is the one-variable case of the generalized clover quiver and is called the \textit{Jordan Quiver}.  In fact, the generating function of this type of quiver is the generating function of the (integer) partitions and  derivation of its asymptotics is a classical problem in analytic combinatorics. The reason that the asymptotic method of this paper does not apply is that the relevant singularities in the one-variable case occur at all possible roots of unity, $1-x^i=0$, and each factor in the product contributes to the asymptotic, while for example in the two-variable case the exponential order of the contribution of $1-x-y=0$ is higher than that of $1-x^2-y^2=0$, etc.

\begin{figure}
\centering
\begin{tikzcd}
\bullet \arrow[out=0,in=90,loop]
\end{tikzcd}
\caption{Jordan Quiver}
\label{jordan}
\end{figure}
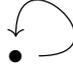

All the oriented cyclic graphs with no multiplicity (multiple edges)
%, forms a class of quivers which we call partition quivers. All the partition quivers 
can be reduced to a quiver consisting of a single vertex and single loop. This loop variable is a
product of edge variables for the cyclic graph. 
%%MCW what does length of the loop mean?
%% SRandAli -- above text is simplified, length removed. 
The simplest example of this class is the two node graph with two oppositely directed edges. The
 asymptotics for this class of quivers follows the asymptotics of partitions.

\subsubsection*{$\bullet$ Bivariate Case}
In the bivariate case ($d=2$), the generating function is 
$$
F(x,y) = \frac{G(x,y)}{H(x,y)}=\prod_{i=1}^\infty (1 - x^i - y^i)^{-1} = \sum_{r,s} a_{rs} x^r y^s.
$$
where $G(x,y)=\prod_{i=2}^{\infty}(1-x^{i}-y^{i})^{-1}$, $H(x,y)=(1-x-y)$.

The asymptotics for $a_{rs}$ as $r+s \to \infty$ and $r/s, s/r$ are bounded away from zero, can be obtained as a special case of the computations above.
Let $\lambda = r/(r+s) \in (0,1)$. This yields the first order asymptotic
$$
a_{rs} \sim \frac{G(\lambda, 1 - \lambda)}{\sqrt{2\pi}}  \frac{(r+s)^{(r+s)}}{r^rs^s}  \sqrt{\frac{r+s}{rs}}.
$$ 
This is uniform in $\lambda$ as long as it stays in a compact subinterval of $(0, 1)$ (alternatively, the slope $r/s$ lies in a compact interval of $(0,\infty)$ --- note that $r/s = \lambda/(1-\lambda)$).
In particular for the main diagonal $r=s$, corresponding to $\lambda = 1/2$, we obtain
$$
a_{nn} \sim \frac{G(1/2,1/2)}{\sqrt{\pi n}}  4^n.$$
The exact value of $G$ at the critical point is not completely explicit, being given by an infinite product. It is a positive real number greater than $1$, since each factor satisfies those same conditions. The minimum value of $G(\lambda)$ over all $\lambda$ occurs when $\lambda = 1/2$ and equals the reciprocal of the \emph{Pochhammer symbol} $(1/2;1/2)_\infty$. This has the approximate numerical value 3.46275. The  value of $G(\lambda)$ approaches $\infty$ as $\lambda\to 0$ or $\lambda\to 1$.

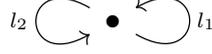
\begin{figure}
\centering
\begin{tikzcd}
\bullet \arrow[out=-30,in=30,loop,swap,"l_1"]
  \arrow[out=150,in=210,loop,swap,"l_2"]
\end{tikzcd}
\caption{Bi-Clover Quiver}
\label{Biclover}
\end{figure}

\subsubsection*{Phase Structure}
\begin{figure}
    \centering
    \includegraphics{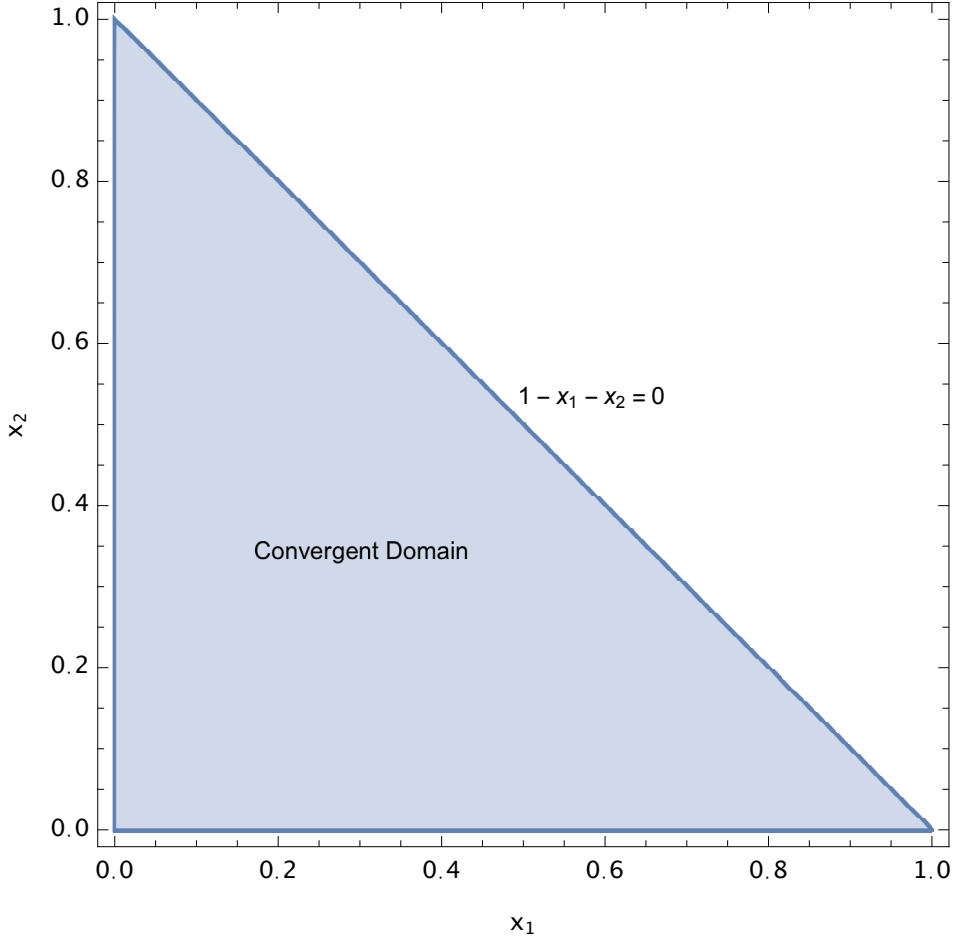}
    \caption{Phase Diagram of Bi-Clover Quiver}
    \label{phase bivariate}
\end{figure}

We start with the simplest example of the class, which is the bivariate clover quiver. Following the discussion in section (3.2), the phase transition line in this example is $1-x_1-x_2=0$ and the phase diagram is depicted in Fig.~\ref{phase bivariate}. In the unrefined case $x_1=x_2$, 
we obtain the critical coupling $\beta^*=\log 2$. In the unrefined case of the generalized clover quiver, we have $\beta^*=\log d$. Similarly the phase diagram of the other examples in this class is a hyper-plane obtained from $H(x_1, ..., x_d)=1-\sum_{j=1}^{d} x_j=0$.
Using Eq.~\eqref{entropy clover}, up to leading order, the entropy and couplings on the critical hypersurface are obtained as
\bea
S^*_{\mathbf{r}} &=& R\log R-\sum_{i=1}^d r_i\log r_i=\sum_{i=1}^d r_i\cdot \beta_i^*,\nonumber\\
\frac{\partial S^*_{\mathbf{r}}}{\partial r_i}&=& \log R- \log r_i =- \log x_i^*= \beta^*_i.
\eea
Notice that critical points obtained as above are the same as the solutions of critical Eqs.~\eqref{Critical Eqs} in the case of generalized clover quivers. The basic physical example in the class of generalized clover quivers is the $\mathbb{C}^3$ quiver gauge theory with three loops in the quiver.

\subsubsection{Oriented Cyclic Quivers with Multiplicity}
An oriented cyclic quiver with multiplicity is an oriented cyclic graph with multiple edges between any two adjacent nodes. The generating functions of these quivers reduce to those of the generalized clover quiver where the number of loops is determined by the number of cycles in the cyclic quiver.
In the following, we present some important examples of oriented cyclic quivers with multiplicity.
\subsubsection*{$\bullet$ Conifold $\mathcal{C}$}
The conifold quiver is an oriented cyclic graph with two nodes and two couples of parallel edges between the nodes. This is a special case of the generalized clover quiver with four variables.
The determinant of the adjacency matrix of the conifold is
\bea
H= 1-l_1-l_2-l_3-l_4,
\eea
where $l_1$, $l_2$, $l_3$, and $l_4$ are product of edge variables in the conifold quiver, see section (2) of \cite{Pa-Ra}.
By direct computation, the critical points and Hessian determinant with the choice of $l_d=l_4$ are 
\bea
l^*_i= \frac{r_i}{R}, \quad \det\mathcal{H}=r_4^{-5}R \prod_{i=1}^{4}r_i.
\eea
where $R=\sum_{i=1}^4 r_i$.
The asymptotic and the dominant terms in entropy can be obtained from the result for the generalized clover quiver,
\bea
a_{\mathbf{r}}\sim\frac{G(\mathbf{l}^*)} {(2\pi)^{3/2}}R^{R+\frac{1}{2}}\prod_{i=1}^{4}r_i^{-r_i-\frac{1}{2}}, \quad S^*_{\mathbf{r}}=-\sum_{i=1}^{4}r_i\log r_i+ R\log R,
\eea 
where $G(\mathbf{l}^*)= \prod_{i=2}^{\infty}(1-l_1^{*i}-l_2^{*i}-l_3^{*i}-l_4^{*i})^{-1}$.
\subsubsection*{$\bullet$ Hirzebruch $F_0$ and del Pezzo $dP_0$ ($\mathbb{C}^3/\mathbb{Z}_3$)}
As other examples of this class we can mention Hirzebruch $F_0$ and del Pezzo $dP_0$, see Fig. 6 (middle) and Fig. 1 in \cite{Fo-Ha}. The generating function of Hirzebruch $F_0$ is the $16$ loop variable case of the generalized clover quiver and del Pezzo $dP_0$  is the generalized clover quiver with $27$ loop variables. Their asymptotics can be obtained as special cases of Eq.~\eqref{free asympt}.

\subsection{Affine $\mathbb{C}^3/\hat{A}_n$ Orbifold Quivers}

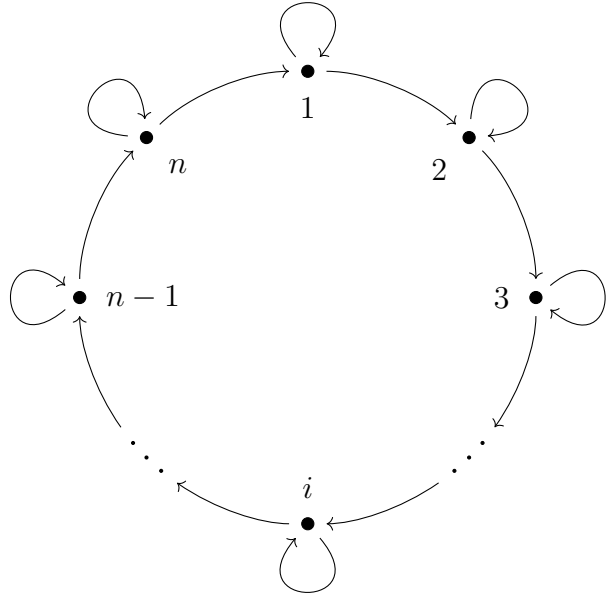
\begin{figure} 
\begin{center}
\begin{tikzpicture}
\foreach \ang\lab\anch in {90/1/north, 45/2/{north east}, 0/3/east, 270/i/south, 180/{n-1}/west, 135/n/{north west}}{
  \draw[fill=black] ($(0,0)+(\ang:3)$) circle (.08);
  \node[anchor=\anch] at ($(0,0)+(\ang:2.8)$) {$\lab$};
  \draw[->,shorten <=7pt, shorten >=7pt] ($(0,0)+(\ang:3)$).. controls +(\ang+40:1.5) and +(\ang-40:1.5) .. ($(0,0)+(\ang:3)$);
}

% Top part of circle, arrows between different nodes and their labels
\foreach \ang\lab in {90/1,45/2,180/{n-1},135/n}{
  \draw[->,shorten <=7pt, shorten >=7pt] ($(0,0)+(\ang:3)$) arc (\ang:\ang-45:3){};
  \node at ($(0,0)+(\ang-22.5:3.5)$) {};
}

% Bottom part of circle, arrows between different nodes and their labels
\draw[->,shorten <=7pt] ($(0,0)+(0:3)$) arc (360:325:3) {};
\draw[->,shorten >=7pt] ($(0,0)+(305:3)$) arc (305:270:3) {};
\draw[->,shorten <=7pt] ($(0,0)+(270:3)$) arc (270:235:3) {};
\draw[->,shorten >=7pt] ($(0,0)+(215:3)$) arc (215:180:3) {};
\node at ($(0,0)+(0-20:3.5)$) {};
\node at ($(0,0)+(315-25:3.5)$) {};
\node at ($(0,0)+(270-20:3.5)$) {};
\node at ($(0,0)+(225-25:3.5)$) {};

% Ellipsis
\foreach \ang in {310,315,320,220,225,230}{
  \draw[fill=black] ($(0,0)+(\ang:3)$) circle (.02);
}
\end{tikzpicture}
\end{center}
\caption{Affine $\mathbb{C}^3/\hat{A}_n$ Orbifold Quivers}
\label{affine quiver}
\end{figure}

The next infinite class of examples consists of the affine $\mathbb{C}^3/\hat{A}_n$ orbifold  quiver theories, see Fig.~\ref{affine quiver}. 
The first observation is that by Eqs.~\eqref{H} and \eqref{loop exp H}, the denominator $H$ for this quiver can be written in terms of the elementary symmetric functions $e_j(x_1, ..., x_n)$,
\bea
H(x_1, ..., x_n, x_c)= -x_c+\sum_{j=0}^{n} (-1)^j e_j(x_1, ..., x_n) = -x_c + \prod_{j=1}^n (1 - x_j).
\eea
Thus the generating function of this quiver can be written as
\bea
F(\mathbf{x})=\sum_{\mathbf{r}} a_{\mathbf{r}} \mathbf{x}^\mathbf{r}=\prod_{i=1}^\infty \Big(-x_c^i+\sum_{j=0}^{n} (-1)^j e_j(x_1^i, ..., x_n^i)\Big)^{-1}.
\eea
First, we choose $g$ function or $x_d$ as the central loop denoted by $x_c$. We denote the other loops in the quiver by $x_i$ for $i=1, ..., n$. We have as above 
$$
G(x_1, \dots, x_n, x_c) = \prod_{k\geq 2} H_k(\mathbf{x}) :=\prod_{i=2}^\infty \left[-x_c^i+\sum_{j=0}^{n} (-1)^j e_j(x_1^i, ..., x_n^i)\right].
$$

Suppose that $r_i/r_c$ and $r_c/r_i$ are fixed and bounded away from zero as $r_i, r_c\rightarrow \infty$.
We claim that the unique critical point and Hessian determinant evaluated at this point are 
\bea
\label{ansatz}
x^*_i= \frac{r_i}{r_c+ r_i}, \quad  x^*_c= \frac{r_c^n}{\prod_{i=1}^{n}(r_c+ r_i)}, \quad \det\mathcal{H}=r_c^{-2n} \prod_{i=1}^{n}r_i (r_i+r_c),
\eea

The proof that this point satisfies $H=0$ follows from the following identity for elementary symmetric function $e_i(r_1, ..., r_n)$,
\bea
\prod_{i=1}^{n}(r_c+r_i)= \sum_{i=0}^{n}r_c^{n-i}e_i(r_1, ..., r_n).
\eea
For the proof of the other critical equations, $\mathbf{r}\times \nabla_{\log} H=0$, first observe that
\bea
\partial_{j} H= \partial_{j} g, \quad \partial_{d} H = -1,
\eea
and  denoting the $g$ function of the quiver with $n$ surrounding loops by $g_n$, we observe that
\bea
\partial_{j} g_n= -g_{n-1}.
\eea
Using the above observation we can prove that the ansatz for $x^*_i$ and $x^*_c$ in Eq.~\eqref{ansatz}, satisfy
\bea
r_c x^*_i \partial_i H= r_i x^*_c \partial_d H.
\eea

The formula for the Hessian determinant in Eq.~\eqref{ansatz} follows from the observation that the Hessian matrix evaluated at the critical point $\mathbf{x}^*$ is diagonal, since
\bea
\frac{x_i x_j}{g^2} \partial_i g \partial_j g= r_i r_j r_c^{-2},\nonumber\\
\frac{x_i x_j}{g} \partial_i \partial_j g= r_i r_j r_c^{-2},
\eea
and moreover, because of the choice $g= x_c$ we have $\partial_i^2 g=0$, and thus by inserting the critical points in Eq.~\eqref{ansatz}, the Hessian matrix, Eq.~\eqref{Hessian 1} becomes
\bea
\mathcal{H}_{ij}= \delta_{ij} (\frac{r_i^2}{r_c^2}+\frac{r_i}{r_c}),
\eea
and therefore the result follows.
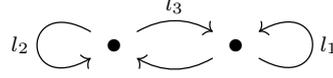
\begin{figure}
\centering
\begin{tikzcd}
\arrow[out=150,in=210,loop,swap,"l_2"]\bullet \arrow[r,bend left,"l_{3}"] & \bullet \arrow[l,bend left,] \arrow[out=-30,in=30,loop,swap,"l_1"]
\end{tikzcd}
\caption{Affine $\mathbb{C}^3/\mathbb{Z}_2$ Quiver}
\label{Affine}
\end{figure}

We now show that the critical point $\mathbf{x}^*$ is strictly minimal and $F = G/H$ has the appropriate form. As in the generalized clover example, this amounts to showing that there are no other zeros of any $H_k$ on the closed polydisk defined by $\mathbf{x}^*$. First note that for each $H_k$, if there is a zero $\mathbf{x}$ then it is a pole  of $1/H_k$, and since $1/H_k$ has nonnegative coefficients \cite{Ma-Ra}, the coordinatewise modulus $\mathbf{x^*}=(|x_1|, \dots, |x_n|, |x_c|)$ is also a pole, so that $|\mathbf{x}|$ is a zero of $H_k$.

We first consider the case $k\geq 2$. Suppose that $0\leq x_i \leq x_i^*$ for all $i$,  that $0\leq x_c \leq x_c^*$. Then if $-x_c^k + \prod_{i=1}^n (1-x_i^k) = 0$, then using the fact that $-x_c^* + \prod_{i=1}^n (1 - x_i^*) = 0$ we obtain
$$
x_c^* - x_c^k = \prod_{i=1}^n (1-x_i^*) - \prod_{i=1}^n (1 - x_i^k).
$$
However $x_c^* - x_c^k > 0$ because $0< x_c \leq x^*_c < 1$, whereas
$$
\prod_{i=1}^n (1-x_i^*) \leq \prod_{i=1}^n (1-x_i^k),
$$
yielding a contradiction. Finally when $k=1$ a similar argument holds: there is no other solution ($\mathbf{x}\neq \mathbf{x^*}$) of $H_1(\mathbf{x})=0$ inside the polydisk. To see this, assume to the contrary that $-x_c + \prod_{i=1}^n (1-x_i) = 0$ and  $-x_c^* + \prod_{i=1}^n (1 - x_i^*) = 0$. This leads to two possibilities: 
\begin{itemize}
\item $x_c - x^*_c = 0$, which yields   $\prod_{i=1}^n (1-x_i^*) = \prod_{i=1}^n (1 - x_i)$ and thus the contradiction $\mathbf{x}=\mathbf{x^*}$;
\item $x_c- x^*_c <0$, which implies $\prod_{i=1}^n (1-x_i^*) - \prod_{i=1}^n (1 - x_i)>0$, again yielding a contradiction because $x_i\leq x_i^*$ for all $i$.
\end{itemize}
\begin{figure}
    \centering
    \includegraphics{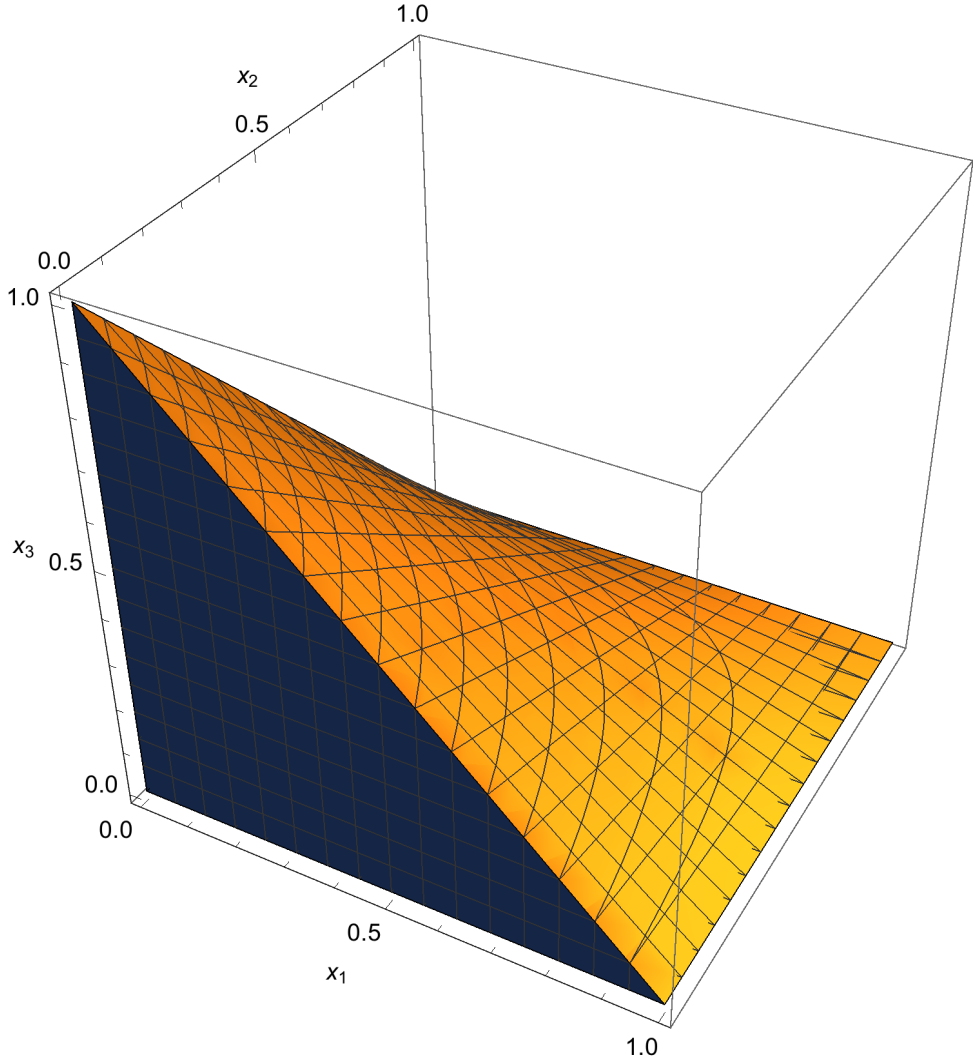}
    \caption{Phase Diagram of $\mathbb{C}^3/\mathbb{Z}_2$. The filled volume which is the convergent domain is separated from the divergent domain by the phase transition hyper-surface.}
    \label{phase affine}
\end{figure}

Hence the asymptotic approximation and the entropy of the general case can be obtained from Eq.~\eqref{asymp form},
\bea
\label{free asympt 3}
a_{\mathbf{r}}&\sim&\frac{G(\mathbf{x}^*)} {(2\pi)^{n/2}}r_c^{-n(r_c+\frac{1}{2})}\prod_{i=1}^{n}r_i^{-r_i-\frac{1}{2}}(r_i+r_c)^{(r_i+r_c+ \frac{1}{2})}, \\ S_{\mathbf{r}}&=& \log G(\mathbf{x}^*)+\sum_{i=1}^{n}\left((-r_i-\frac{1}{2})\log r_i+(r_i+r_c+\frac{1}{2})\log (r_i+r_c)\right)- n (r_c+\frac{1}{2})\log r_c, \nonumber
\eea 
where $G(\mathbf{x}^*)= \prod_{i=2}^\infty H(x_1^{*i}, x_2^{*i}, ..., x_n^{*i}, x_c^{*i})^{-1}$.

\subsubsection*{Phase Structure}
We start with the simplest example of the class, which is affine $\mathbb{C}^3/\mathbb{Z}_2$, see Fig.~\ref{Affine}. 
As we discussed in section (3.2), the phase transition hyper-surface in this example is $1-x_1-x_2-x_3+x_1 x_2=0$ and the phase diagram is depicted in Fig.~\ref{phase affine}. In the general case, the phase diagram of this class is a hyper-surface obtained from $H(x_1, ..., x_n, x_c)=0$ which implies $ x_c=\sum_{j=0}^{n} (-1)^j e_j(x_1, ..., x_n)$. 
From Eq.~\eqref{free asympt 3} and  Eq.~\eqref{ansatz}, the critical relation on the phase transition hyper-surface, up to leading order, for the affine $\mathbb{C}^3/\hat{A}_n$, 
yields
\bea
S^*_{\mathbf{r}} &=& \sum_{i=1}^{n}\Big(-r_i\log r_i+(r_i+r_c)\log (r_i+r_c)\Big)- n r_c\log r_c=\sum_{i=1}^n r_i\cdot \beta_i^*+ r_c\cdot \beta_c^*,\nonumber\\
\frac{\partial S^*_{\mathbf{r}}}{\partial r_i}&=& \log r_i- \log (r_i+r_c) =- \log x_i^*= \beta^*_i,\ \text{for} \ i=1, 2, ..., n,\nonumber\\
\frac{\partial S^*_{\mathbf{r}}}{\partial r_c}&=& n\log r_c- \sum_{i=1}^n\log (r_i+r_c) =- \log x_c^*= \beta^*_c.
\eea
Notice that critical points obtained as above are the same as the solutions of critical equations Eqs.~\eqref{Critical Eqs}, presented in Eq.~\eqref{ansatz}.

\section{Conclusions and Forthcoming Research}
\label{sec:future}

In this work
%we introduced a new method from analytic combinatorics to study the asymptotic limit of the gauge invariant chiral operators in quiver gauge theories. 
we adapted recent results in the multivariate asymptotic analysis of generating functions to study the asymptotic counting of operators in the chiral ring of free $\mathcal{N} =1$ quiver gauge theories. 
 We obtained an explicit asymptotic formula for two infinite classes of examples. A next step is to consider other classes of examples, e.g. infinite classes of orbifolds such as $\mathbb{C}^3/\mathbb{Z}_n$ and $\mathbb{C}^3/A_n$, ADE class, $L^{a,b,c}$, $Y^{p,q}$ Sasaki-Einstein spaces. A general formula for the asymptotics for general quivers, expressed in terms of the weighted adjacency matrix, would be an interesting goal.

%In this study, we did not discuss the quivers with multiplicity, i.e. multiple edges between two nodes, except the Kronecker quiver which is reducible to the non-multiple generalized clover quiver. The (irreducible) multiple quivers can be considered with similar methods to those used in this article, and their asymptotic results  can be obtained as a straightforward generalization of the current results.  

In this work, we only considered the first (leading) asymptotic term in the asymptotic series. The current results  in analytic combinatorics for the asymptotic counting of multivariate generating functions \cite{Ra-Wi} allow for  the computation of 
sub-leading contributions in the asymptotics. These results should be applied to 
obtain the higher order asymptotics in quiver gauge theories.

In this work, the matter content of the gauge theory is restricted to matter in bifundamental representations.  One can also consider fundamental matter and quiver gauge theories with flavours. The generating functions for counting gauge invariant operators in quivers with flavours are  obtained in \cite{Ma-Ra}. Thus it should be straightforward to generalize the method of this work to quivers with flavours and obtain the asymptotic counting of chiral operators in these  theories.

In this paper, we considered the zero super-potential limit and exploited the availability of  general formula for the generating functions of the multi-trace chiral operators. However, as we mentioned, the $W\neq 0$ sector of the quiver gauge theories and the counting of the chiral operators have been studied vastly. Generating function technology
% , such as Hilbert series, plethystic exponential/logarithm 
has been  introduced and applied successfully to very general classes of quivers \cite{Be-Fe-Ha-He,Fe-Ha-He,Do}. 
 The multivariate asymptotic analysis employed here would be useful in obtaining  results 
 for asymptotics of counting in general quiver gauge theories.

The counting of states for the bi-clover quiver (Section~\ref{sec:classes}) is the two-matrix counting which has 
been recently discussed in the context of small black holes in AdS/CFT \cite{Ber2,Ber1}. 
We expect that the consideration of higher order asymptotics will be relevant to this discussion.
It would also be interesting to explore the more general quiver asymptotics in the context of AdS/CFT. 
%In this work, we mostly discussed the mathematical aspects of the asymptotic counting of the holomorphic gauge invariant chiral operators in free quiver theories. Certainly, more physics oriented discussions, towards the physical applications of the obtained results in terms of the entropy and phase structure of the quiver gauge theories and their implications for the full quiver gauge theory and black holes, are highly desirable.

\section{Acknowledgement}

The research of S.R. is supported by the STFC consolidated grant ST/L000415/1 ``String Theory, Gauge Theory \& Duality'' and  a Visiting Professorship at the University of the Witwatersrand, funded by a Simons Foundation grant to the Mandelstam Institute for Theoretical Physics. 
A.Z. is supported by research funds from the Centre for Research in String Theory, School of Physics and Astronomy, Queen Mary University of London, and National Institute for Theoretical Physics, School of Physics and Mandelstam Institute for Theoretical Physics, University of the Witwatersrand. We are deeply grateful to Robert de Mello Koch for valuable discussions and collaboration in the early stages of this work.

\appendix
\section{Hessian Determinant of Generalized Clover Quiver} 
\label{App A}

$\mathfrak{H}$ is an $n \times n $ matrix where $n = d-1$. 
Consider the matrix $\mathfrak{H}$ which is a rescaled version of $ \cH$. 
\bea\label{Hdef}  
\mathfrak{H}_{ ij} = { x_i x_j } +  { x_i \delta_{ ij} }.  
\eea
The indices  $i,j$ take values from $ 1$ to $n-1$. 
We have 
\bea\label{detFormula}  
\det \mathfrak{H} = \epsilon_{ i_1 \cdots i_n } \mathfrak{H}_{ 1 i_1 } \mathfrak{H}_{ 2 i_2} \cdots \mathfrak{H}_{ n i_n }. 
\eea
Define the  matrix $\mathfrak{X}$ which  is $\diag ( x_1 , x_2 , \cdots , x_n )$. 
Equivalently
\bea 
\mathfrak{X}_{ i j } = x_i \delta_{ ij} .
\eea
Using the expression \eqref{Hdef} in \eqref{detFormula}, we encounter  terms 
where all the $\mathfrak{H}$ factors contribute the second term, linear in $x$. 
These terms are 
\bea 
&& \epsilon_{ i_1 \cdots i_n } x_1 \delta_{ 1 i_1} x_2 \delta_{ 2 i_2} \cdots 
  x_n \delta_{ n i_n } \cr 
   && =\det \mathfrak{X}.
\eea
Next there are terms where we pick, from one of the $\mathfrak{H}$'s in \eqref{detFormula}
the  term quadratic in $x$'s, and from the rest the linear term. 
Taking the distinguished $\mathfrak{H}$  to be the first one, we have 
\bea 
&& \epsilon_{ i_1 \cdots i_n }  x_1 x_{i_1} ~  x_2 \delta_{ 2 i_2} ~ x_3 \delta_{ 3 i_3} ~ \cdots 
 ~ x_n \delta_{ n i_n }  \cr 
&& = (\det \mathfrak{X}) x_{ i_1} \epsilon_{ i_1 , 2 , \cdots , n }  \cr 
&& = ( \det \mathfrak{X} ) x_1. 
\eea 
Summing over all the possible choices of the $\mathfrak{H}$ factor contributing the quadratic term 
in $x$'s, we have 
\bea 
( \det \mathfrak{X} ) ( \tr \mathfrak{X} ).  
\eea
We also have to consider terms where we pick quadratic terms from two or more $\mathfrak{H}$'s. 
These lead to terms of the form 
\bea 
\epsilon_{ i_1  \cdots  i_n } x_{ i_1} x_{ i_2} \cdots 
\eea
which vanish due to the symmetry of the $x$'s and the antisymmetry of the $ \epsilon$. 
We therefore conclude 
\bea 
\det \mathfrak{H}  = (\det \mathfrak{X})  (1+\tr \mathfrak{X}).
\eea
Then, it is straightforward to obtain Eq. \eqref{Clover det}.

%% no longer needed
\if01
\section{Dominance of $H(\textbf{x})=0$ in the Asymptotics}
\label{App B}
In this part, we move towards analytic proof of the dominancy of lowest power term (in the infinite product generating function) in the asymptotic analysis. In the generalized clover quiver, we prove that the leading contribution to the asymptotics is from the solution of $H(\textbf{x})=0$ and solutions of the higher terms $H(\textbf{x}^k)=0$ for $k>1$ are subleading in the asymptotics. In the affine orbifold quiver, based on analytic computations and numerical evidences, we claim $H=0$ is the dominant term in the asymptotics. Our strategy for the proof is to show that the solution of $H(\textbf{x})=0$ is the closet solution to the origin and it is the first encountered singularity. Another crucial point is that the critical points (solution of $H(\textbf{x})=0$) are not the singularities of $G(\textbf{x})$ and for the inverse function of $G$ evaluated at critical points we have $G^{-1}(\textbf{x})>0$.

%In the generalized clover quiver we want to show that the first encountered singularity is from the solution of $H=1-\sum_{i=1}^d x_i =0$...
\subsubsection*{Generalized clover quiver}
First, we consider the generalized clover quiver with $H_k(\textbf{x})= 1-\sum_{j=1}^{d} x_j^k$. First, we check that the zero of $H_1(\textbf{x})=0$ denoted by $\textbf{x}^{(1)}$ satisfying $H_1(\textbf{x}^{(1)})=0$ are not zeros of the higher terms with $k>1$, i.e. $H_k(\textbf{x}^{(1)})\neq 0$.

Proof. Using $H_1(\textbf{x}^{(1)})=0$, for any $k$ we have 
\bea
 \left(\sum_{i=1}^d x_i^{(1)}\right)^k=1.
\eea
and this leads to
\bea
H_k(\textbf{x}^{(1)})= 1-\sum_{i=1}^d (x_i^{(1)})^k=\left(\sum_{i=1}^d x_i^{(1)}\right)^k- \sum_{i=1}^d (x_i^{(1)})^k>0.
\eea
Moreover, using the solutions of the critical equations $x_i= \frac{r_i}{R}$, or assuming $x_i=e^{-g_i}<1$, we have
\bea
H_{k+1}(\textbf{x}^{(1)})- H_k(\textbf{x}^{(1)})= \sum_{i=1}^d ((x_i^{(1)})^n - (x_i^{(1)})^{n+1}) >0.
\eea
Using $G(\mathbf{x})=\prod_{i=2}^\infty H_i(\textbf{x})^{-1}$, we obtain $G^{-1}(\textbf{x}^{(1)})>0$. Thus, $\textbf{x}^{(1)}$ are not the solution of $G^{-1}(\textbf{x})=0$.

Next, we show that the solutions of higher terms $H_k(\textbf{x}^{(k)})=0$ for arbitrary $k>1$ are further from the origin compare to solutions $\textbf{x}^{(1)}$ satisfying $H_1(\textbf{x}^{(1)})=0$. In other words we have $|\textbf{x}^{(k)}|>|\textbf{x}^{(1)}|$ where $\left(\textbf{x}^{(j)}\right)^2= \sum_{i=1}^{d}\left(x_i^{(j)}\right)^2$, for any $j$ from $1$ to $\infty$ including $k$, and then consider a set of straight lines $x_d^{(j)}=m_i x_i^{(j)}$ for $i=1$ to $d-1$ (there is no sum over repeated indices). Then we have
\bea
\left(\textbf{x}^{(j)}\right)^2 = \left(x_d^{(j)}\right)^2\left(1+\sum_{i=1}^{d-1}\frac{1}{m_i^2}\right).
\eea
Thus to show $|\textbf{x}^{(k)}|>|\textbf{x}^{(1)}|$ we need to show $x_d^{(k)}>x_d^{(1)}$. From the solutions of $H_1(\textbf{x}^{(1)})=0$ and $H_k(\textbf{x}^{(k)})=0$ we have 
\bea
\left(x_d^{(1)}\right)^k=\left(1+\sum_{i=1}^{d-1}\frac{1}{m_i}\right)^{-k},\quad \left(x_d^{(k)}\right)^k=\left(1+\sum_{i=1}^{d-1}\frac{1}{m_i^k}\right)^{-1}.
\eea
Using 
\bea
\left(1+\sum_{i=1}^{d-1}\frac{1}{m_i}\right)^{k}> \left(1+\sum_{i=1}^{d-1}\frac{1}{m_i^k}\right),
\eea
one can show that $x_d^{(k)}> x_d^{(1)}$ and this proves $|\textbf{x}^{(k)}|>|\textbf{x}^{(1)}|$. A stronger result but not necessary for our work is $|\textbf{x}^{(k+1)}|>|\textbf{x}^{(k)}|$ and for this we need to prove 
\bea
\left(1+\sum_{i=1}^{d-1}\frac{1}{m_i^k}\right)^{k+1}> \left(1+\sum_{i=1}^{d-1}\frac{1}{m_i^{k+1}}\right)^k.
\eea

In the following, we prove that in the bi-clover quiver, the first singularities that are encountered are those which are the solutions $x_1^{(1)}$, $x_2^{(1)}$ satisfying $H_1(x_1^{(1)}, x_2^{(1)})=0$. As we get to higher powers and consider the solutions $x_1^{(n)}$, $x_2^{(n)}$ satisfying $H_n(x_1^{(n)}, x_2^{(n)})=0$, the singularities get further from the origin.

The solution of $H_n(x_1, x_2)=0$, denoted by $x_1^{(n)}$, $x_2^{(n)}$ parameterize a curve. The distance of any point (parameterized by $m$) on this curve can be computed by $r_n\coloneqq((x_1^{(n)})^2+(x_2^{(n)})^2)^{1/2}$ where $x_1^{(n)}$ and $x_2^{(n)}$ are the solutions of the following set of equations,
\bea
(x_1^{(n)})^n+(x_2^{(n)})^n=1, \quad x_2^{(n)}= m x_1^{(n)}.
\eea
Thus, we have 
\bea
r_n= (1+m^2) (\frac{1}{1+m^n})^{2/n}.
\eea
We want to show $r_{n+1}>r_n$. To prove this, we define $s_n\coloneqq (1+m^n)^{2/n}$ and now we should prove $s_n>s_{n+1}$. To simplify more we can consider $t_n\coloneqq s_n^{n(n+1)/2}=(1+m^n)^{n+1}$ and $t_{n+1}\coloneqq s_{n+1}^{n(n+1)/2}= (1+m^{n+1})^n$ and to show that $r_n$ is increasing with $n$, we need to show $t_n-t_{n+1}>0$.

Expanding $t_n$ and $t_{n+1}$, we have
\bea
t_n&=&\sum_{k=0}^{n+1} \binom{n+1}{k} m^{nk}=1+(n+1) m^n + \binom{n+1}{2} m^{2n}+...,\nonumber\\
t_{n+1}&=&\sum_{k=0}^{n} \binom{n}{k} m^{(n+1)k}= 1+ n m^{n+1} + \binom{n}{2} m^{2n+2}+ ... .
\eea
There are two cases. First consider $m<1$, and now we have
\bea
t_n-t_{n+1}&=&\sum_{k=0}^n m^{nk}\bigg( \binom{n+1}{k}- \binom{n}{k} m^k \bigg)\nonumber\\
&=& \sum_{k=0}^n m^{nk}\bigg(\frac{1}{k}\binom{n}{k-1}\Big(n (1-m^k)+\big(1+m^k(k-1)\big)\Big)\bigg),
\eea
which is apparently positive.

For $m>1$, we can write 
\bea
t_n = m^{n(n+1)}(1+m^{-n})^{n+1}, \quad t_{n+1} = m^{n(n+1)}(1+m^{-n-1})^n.
\eea
Similarly, expanding $t_n$ and $t_{n+1}$ and computing the difference, we obtain
\bea
t_n-t_{n+1}&=& \sum_{k=0}^n m^{-kn+n(n+1)}\bigg(\binom{n+1}{k}- \binom{n}{k} m^{-k}\bigg)\nonumber\\
&=& \sum_{k=0}^n m^{-kn+n(n+1)}\Bigg(\frac{1}{k}\binom{n}{k-1}\bigg(n (1-m^{-k})+\left(1+m^{-k}(k-1)\right)\bigg)\Bigg),\nonumber\\
\eea
which proves that the difference is positive in this case.

\subsubsection*{Affine orbifold $\mathbb{C}^3/\hat{\mathbb{A}}_n$ quiver}
Next, we consider the affine orbifold $\mathbb{C}^3/\hat{\mathbb{A}}_n$ quiver with
\bea
H_k(x_1, ..., x_n, x_c)= -x_c^k+\sum_{i=0}^{n} (-1)^i e_i(x_1^k, ..., x_n^k).
\eea
First, we check that the zero of $H_1(\textbf{x})=0$ denoted by $\textbf{x}^{(1)}$ satisfying $H_1(\textbf{x}^{(1)})=0$ are not zeros of the higher terms with $k>1$, i.e. $H_k(\textbf{x}^{(1)})\neq 0$. Using the following identity
\bea
\label{elem iden}
\prod_{i=1}^{n}(1-x_i)= \sum_{i=0}^{n}(-1)^{i}e_i(x_1, ..., x_n),
\eea
we see that solutions $\textbf{x}^{(1)}$ of $H_1(\textbf{x}^{(1)})=0$ satisfy
\bea
\label{(1) iden}
x_c^{(1)}=\prod_{i=1}^{n}(1-x_i^{(1)}).
\eea
Now, we compute an arbitrary higher term,
\bea
H_m(\textbf{x}^{(1)})= -(x_c^{(1)})^m+\prod_{i=1}^{n}(1-(x_i^{(1)})^m),
\eea
and using Eq.~\eqref{(1) iden}, we obtain
\bea
H_m(\textbf{x}^{(1)})= -\prod_{i=1}^{n}(1-x_i^{(1)})^m+\prod_{i=1}^{n}(1-(x_i^{(1)})^m).
\eea
%Using the solutions of critical equations $x_i= \frac{r_i}{r_i+r_c}<1$, and expanding and keeping the dominant term  we find
%\bea
%H(x^m)\sim m x_i >0.
%\eea
%This leads to  
%\bea
%H(x^{m+1})- H(x^m)\sim x_i >0.
%\eea
%Now, we can directly compute 
%\bea
%H(x^{m+1})- H(x^m)= x_c^m- x_c^{m+1} + \prod_{i=1}^{n}(1-x_i^{m+1}) -\prod_{i=1}^{n}(1-x_i^m).
%\eea
%First, from the critical equation we have
%\bea
%x_c^m- x_c^{m+1}= \prod_{i=1}^{n}(1-x_i)^m - \prod_{i=1}^{n}(1-x_i)^{m+1}.
%\eea
%Then, expanding and keeping the dominant term using $x_i<1$, we find 
%\bea
%H(x^{m+1})- H(x^m)\sim x_i >0.
%\eea
For $\mathbb{C}^3/\mathbb{Z}_2$ we have
\bea
H_1(\textbf{x}^{(1)})=1-x_1^{(1)}-x_2^{(1)}-x_3^{(1)} +x_1^{(1)} x_2^{(1)}=0.
\eea
The next higher order term is
\bea
H_2(\textbf{x}^{(1)})=1-(x_1^{(1)})^2-(x_2^{(1)})^2-(x_3^{(1)})^2 +(x_1^{(1)})^2 (x_2^{(1)})^2.
\eea
Using $(x_1^{(1)}+x_2^{(1)}+x_3^{(1)} -x_1^{(1)} x_2^{(1)})^2=1$,
we find 
\bea
H_2(\textbf{x}^{(1)})=1-(x_1^{(1)})^2-(x_2^{(1)})^2-(x_3^{(1)})^2 +(x_1^{(1)})^2 (x_2^{(1)})^2=1+ 2 x_3^{(1)}(x_1^{(1)}+x_2^{(1)})>0.
\eea
We have numerical evidences that for other examples in this class of quivers, the similar results hold.

Consider the solutions $\textbf{x}^{(1)}$ satisfying $H_1(\textbf{x}^{(1)})=0$ and $\textbf{x}^{(k)}$ satisfying $H_k(\textbf{x}^{(k)})=0$, the goal is to show $|\textbf{x}^{(k)}|>|\textbf{x}^{(1)}|$. Using $x_c^{(j)}=m_i x_i^{(j)}$ and identity \eqref{elem iden}
we obtain
\bea
\label{aff orb crit}
\left(x_c^{(1)}\right)= \prod_{i=1}^{n}\left(1-\frac{x_c^{(1)}}{m_i}\right), \quad \left(x_c^{(k)}\right)^k= \prod_{i=1}^{n}\left(1-\left(\frac{x_c^{(k)}}{m_i}\right)^k\right).
\eea
Similar to clover case and with the choice $x_c=x_d$ we need to show that $x_c^{(k)}>x_c^{(1)}$ and using above Eq.~\eqref{aff orb crit}, we must have
\bea
\prod_{i=1}^{n}\left(1-\frac{x_c^{(1)}}{m_i}\right)^k <\prod_{i=1}^{n}\left(1-\left(\frac{x_c^{(k)}}{m_i}\right)^k\right).
\eea
Assume there exists a solution of (L.H.S) equation in Eq.~\eqref{aff orb crit}, $x_c^{(1)}=f(m_1, m_2, ..., m_n)$, then the solution of (R.H.S) equation is $x_c^{(k)}=(f(m_1^k, m_2^k, ..., m_n^k))^{1/k}$. There are numerical evidences in many examples that $x_c^{(k)}$ and $x_c^{(1)}$ as solutions of Eqs.~\eqref{aff orb crit} satisfy $x_c^{(k)}>x_c^{(1)}$.
Finally, in general case for any quiver, based on numerical evidences, we expect that first singularity and the dominant term is $H=0$. 
%Use the loop expansion...

\fi

\bibliographystyle{plain}
\bibliography{references}

\begin{thebibliography}{10}

\bibitem{Aha}
Ofer Aharony, Joseph Marsano, Shiraz Minwalla, Kyriakos Papadodimas, and Mark
  Van~Raamsdonk.
\newblock The {H}agedorn/deconfinement phase transition in weakly coupled large
  ${N}$ gauge theories.
\newblock In {\em Lie Theory And Its Applications In Physics V}, pages
  161--203. World Scientific, 2004.

\bibitem{Be-Fe-Ha-He}
Sergio Benvenuti, Bo~Feng, Amihay Hanany, and Yang-Hui He.
\newblock Counting {BPS} operators in gauge theories: quivers, syzygies and
  plethystics.
\newblock {\em Journal of High Energy Physics}, 2007(11):050, 2007.

\bibitem{Ber2}
David Berenstein.
\newblock Negative specific heat from non-planar interactions and small black
  holes in ads/cft.
\newblock {\em arXiv preprint arXiv:1810.07267}, 2018.

\bibitem{Ber1}
David Berenstein.
\newblock Submatrix deconfinement and small black holes in ads.
\newblock {\em arXiv preprint arXiv:1806.05729}, 2018.

\bibitem{CDSW}
Freddy Cachazo, Michael~R. Douglas, Nathan Seiberg, and Edward Witten.
\newblock {Chiral rings and anomalies in supersymmetric gauge theory}.
\newblock {\em Journal of High Energy Physics}, 12:071, 2002.

\bibitem{Cve}
Drago\v{s}~M. Cvetkovi\'{c}, Michael Doob, and Horst Sachs.
\newblock {\em Spectra of Graphs: Theory and Application}.
\newblock Academic Press, New York, 1980.

\bibitem{Do}
FA~Dolan.
\newblock Counting {BPS} operators in $\mathcal{N}=4$ {SYM}.
\newblock {\em Nuclear Physics B}, 790(3):432--464, 2008.

\bibitem{Do-Mo}
Michael~R Douglas and Gregory Moore.
\newblock D-branes, quivers, and {ALE} instantons.
\newblock {\em arXiv hep-th/9603167}, 1996.

\bibitem{Fe-Ha-He}
Bo~Feng, Amihay Hanany, and Yang-Hui He.
\newblock Counting gauge invariants: the plethystic program.
\newblock {\em Journal of High Energy Physics}, 2007(03):090, 2007.

\bibitem{Fo-Ha}
Davide Forcella, Amihay Hanany, Yang-Hui He, and Alberto Zaffaroni.
\newblock The master space of $\mathcal{N}=1$ gauge theories.
\newblock {\em Journal of High Energy Physics}, 2008(08):012, 2008.

\bibitem{Franco:2005rj}
Sebastian Franco, Amihay Hanany, Kristian~D. Kennaway, David Vegh, and Brian
  Wecht.
\newblock {Brane dimers and quiver gauge theories}.
\newblock {\em Journal of High Energy Physics}, 01:096, 2006.

\bibitem{Ful}
William Fulton.
\newblock {\em Introduction to Toric Varieties. (AM-131)}.
\newblock Princeton University Press, 1993.

\bibitem{Gr-Wi-Sc}
M.~B. Green, J.~H. Schwarz, and E.~Witten.
\newblock {\em Superstring Theory}.
\newblock Cambridge University Press, 1987.

\bibitem{Hanany:2005ve}
Amihay Hanany and Kristian~D. Kennaway.
\newblock {Dimer models and toric diagrams}.
\newblock {\em arXiv hep-th/0503149}, 2005.

\bibitem{Ha-Za}
Amihay Hanany and Alberto Zaffaroni.
\newblock The master space of supersymmetric gauge theories.
\newblock {\em Advances in High Energy Physics}, 2010, 2010.

\bibitem{Ken}
Kristian~D Kennaway.
\newblock Brane tilings.
\newblock {\em International Journal of Modern Physics A}, 22(18):2977--3038,
  2007.

\bibitem{Lucietti:2008cv}
James Lucietti and Mukund Rangamani.
\newblock {Asymptotic counting of BPS operators in superconformal field
  theories}.
\newblock {\em J. Math. Phys.}, 49:082301, 2008.

\bibitem{Malda1997}
Juan~Martin Maldacena.
\newblock {The Large-N limit of superconformal field theories and
  supergravity}.
\newblock {\em Int. J. Theor. Phys.}, 38:1113--1133, 1999.

\bibitem{Ma-Ra}
Paolo Mattioli and Sanjaye Ramgoolam.
\newblock Quivers, words and fundamentals.
\newblock {\em Journal of High Energy Physics}, 2015(3):105, 2015.

\bibitem{McG}
James McGrane, Sanjaye Ramgoolam, and Brian Wecht.
\newblock Chiral ring generating functions \& branches of moduli space.
\newblock {\em arXiv preprint arXiv:1507.08488}, 2015.

\bibitem{Pa-Ra}
Jurgis Pasukonis and Sanjaye Ramgoolam.
\newblock Quivers as calculators: counting, correlators and {R}iemann surfaces.
\newblock {\em Journal of High Energy Physics}, 2013(4):94, 2013.

\bibitem{Pe}
Robin Pemantle.
\newblock Analytic combinatorics in $d$ variables: an overview.
\newblock {\em Algorithmic probability and combinatorics}, 520:195--220, 2010.

\bibitem{Pe-Wi}
Robin Pemantle and Mark~C. Wilson.
\newblock Twenty combinatorial examples of asymptotics derived from
  multivariate generating functions.
\newblock {\em SIAM Review}, 50(2):199--272, 2008.

\bibitem{Pe-Wi-book}
Robin Pemantle and Mark~C. Wilson.
\newblock {\em Analytic combinatorics in several variables}, volume 140.
\newblock Cambridge University Press, 2013.

\bibitem{Ra-Wi}
Alexander Raichev and Mark~C. Wilson.
\newblock Asymptotics of coefficients of multivariate generating functions:
  improvements for smooth points.
\newblock {\em Electronic Journal of Combinatorics}, 15(1):89, 2008.

\bibitem{Wit1}
Edward Witten.
\newblock Anti de {S}itter space and holography.
\newblock {\em arXiv preprint hep-th/9802150}, 1998.

\end{thebibliography}

\end{document}